\documentclass[11pt]{article}
\usepackage{amsfonts}
\usepackage{latexsym}
\usepackage{amsmath,amssymb}
\usepackage{verbatim}
\usepackage{setspace}
\usepackage{color}
\usepackage{tikz}
\usepackage{mathdots}
\usepackage{yhmath}
\usepackage{cancel}
\usepackage{color}
\usepackage{mathrsfs}  
\usepackage{array}
\usepackage{multirow}
\usepackage{amssymb}
\usepackage{tabularx}
\usepackage{hyperref}
\usepackage{booktabs}
\numberwithin{figure}{section}

\usepackage[textheight=9in, textwidth=6.5in, letterpaper]{geometry}
\numberwithin{equation}{section}

   \makeatletter
  \let\over=\@@over \let\overwithdelims=\@@overwithdelims
  \let\atop=\@@atop \let\atopwithdelims=\@@atopwithdelims
  \let\above=\@@above \let\abovewithdelims=\@@abovewithdelims
\renewcommand\section{\@startsection {section}{1}{\z@}%
                                   {-3.5ex \@plus -1ex \@minus -.2ex}%nn
                                   {2.3ex \@plus.2ex}%
                                   {\normalfont\large\bfseries}}

\renewcommand\subsection{\@startsection{subsection}{2}{\z@}%
                                     {-3.25ex\@plus -1ex \@minus -.2ex}%
                                     {1.5ex \@plus .2ex}%
                                     {\normalfont\bfseries}}

\begin{document}
%\begin{titlepage}
\unitlength = 1mm
\ \\
\vskip1cm
\begin{center}

{ \LARGE {\textsc{Collinear Corrections to the Cachazo-Strominger Soft Theorem}}}

\vspace{0.8cm}
Eivind J\o rstad and Sruthi A. Narayanan

\vspace{1cm}

{\it  Perimeter Institute for Theoretical Physics,\\
31 Caroline Street North, Waterloo, Ontario, Canada N2L 2Y5}

\begin{abstract}
 Soft theorems describe the behavior of scattering amplitudes when one or several external particles are taken to be energetically soft. In tree-level gravity there are universal soft theorems for the three leading orders in the soft expansion, and they can be shown to be equivalent to Ward identities of asymptotic symmetries. While the leading and subleading symmetries are understood as supertranslations and superrotations respectively, the precise symmetry interpretation of the sub-subleading soft theorem is still a matter of investigation. The form of the sub-subleading soft graviton theorem was elucidated by Cachazo and Strominger using a BCFW expansion of graviton amplitudes. In this work we show that consistency with results based on asymptotic charges requires a careful treatment of collinear singularities in the amplitude, giving rise to collinear corrections to the usual Cachazo-Strominger soft theorem. 
 \end{abstract}
\vspace{0.5cm}

\vspace{1.0cm}

\end{center}

%\end{titlepage}S3
\pagestyle{empty}
\pagestyle{plain}
\newpage
\tableofcontents

\pagenumbering{arabic}

\section{Introduction}

Originating with the work of Weinberg in 1965~\cite{Weinberg:1965nx}, the study of soft theorems has shown to be tremendously useful in understanding the structure of amplitudes \cite{Bern:1998sv,Zhou:2022orv}. For example, one can use the soft theorems to recursively build up higher point amplitudes. Soft theorems describe the behavior of gauge theory and gravity amplitudes when the energy of one or more external particles are taken to be energetically soft. At tree-level, low order soft theorems are universal\footnote{By universal, we mean that the soft factorization is independent of the content of the theory. This is only true at finitely many low orders of the soft limit at tree-level. For example, beyond sub-subleading level in gravity it is expected that the tree-level soft theorem contains a universal part~\cite{Hamada:2018vrw} and a non-universal part, the former of which should be related to a symmetry law. }and have been shown to be equivalent to Ward identities of asymptotic symmetries~\cite{ Hamada:2018vrw,Strominger:2017zoo}. Asymptotic symmetries have, in various forms, played an important role in our understanding of a theory of quantum gravity. As such, defining the full set of symmetries that correspond to each order of the soft limit is a problem that has received much recent attention~\cite{Strominger:2021mtt}.

While studying the implications of asymptotic symmetries for a theory of quantum gravity, the field of ``celestial holography" was born. Celestial holography~\cite{Pasterski:2016qvg} proposes a putative duality between a bulk theory of quantum gravity in $d$ dimensions and a ``celestial" conformal field theory that lives on the $(d-2)$-dimensional boundary celestial sphere. Given that the bulk $d$-dimensional Lorentz symmetry is manifest as $(d-2)$-dimensional conformal symmetry on the boundary sphere, one can recast bulk $\mathcal{S}$-matrix elements as boundary conformal correlators~\cite{Pasterski:2016qvg}. With this prescription, the aforementioned soft theorems in the bulk correspond to the insertion of associated conserved currents in the boundary correlators. The associated Ward identities are also referred to as conformally soft theorems~\cite{Donnay:2018neh,Pate:2019mfs,Puhm:2019zbl}. See~\cite{Raclariu:2021zjz,Pasterski:2021rjz} for detailed reviews. Among other aspects, the algebra~\cite{Guevara:2021abz} of these conserved currents has been a recent focus of study \cite{Melton:2024jyq, Ball:2024oqa, Cresto:2024fhd,Cresto:2025bfo} because it is a natural structure to study in a conformal field theory. However, given that the operators and their correlation functions can be directly mapped to bulk fields and scattering amplitudes, these algebraic properties should be manifest in the study of amplitudes alone.

In the last few years, there has been significant progress in this direction by understanding the origin of the conserved currents from a canonical bulk perspective \cite{Freidel:2021dfs,Freidel:2021ytz,Freidel:2023gue}. In particular, since the soft theorems are Ward identities of asymptotic symmetries, there should be associated Noether charges that can be constructed in the bulk whose Poisson brackets admit the same algebraic structure as that observed in the boundary CFT. The details of this construction can be found in the aforementioned papers but the basic logic is as follows: We denote the total conserved charge as $\mathcal{Q}^\pm$ where the $\pm$ denotes the action on incoming or outgoing states. In terms of the $\mathcal{S}$-matrix, the conservation law is written as~\cite{Strominger:2017zoo} 
\begin{equation}
\langle\rm{out}|\mathcal{Q}^+\mathcal{S}-\mathcal{S}\mathcal{Q}^-|\rm{in}\rangle = 0.
\end{equation}
These charges can be split into a hard part and a soft part, usually denoted by $\mathcal{Q}^\pm = \mathcal{Q}^\pm_H+\mathcal{Q}^\pm_S$. The commutator of the soft charge $\mathcal{Q}^\pm_S$ with the $\mathcal{S}$-matrix is realized as the insertion of a soft particle, like a graviton, into the $\mathcal{S}$-matrix. The commutator of the hard charge $\mathcal{Q}^\pm_H$ with the $\mathcal{S}$-matrix is realized as generating the action of the associated symmetry on the hard matter. Heuristically, one obtains the following
\begin{equation}
\langle{\rm{out}}|\mathcal{Q}_S^+\mathcal{S}-\mathcal{S}\mathcal{Q}_S^-|{\rm{in}}\rangle = \langle{\rm{out}}|\mathcal{Q}^+_H\mathcal{S}-\mathcal{S}\mathcal{Q}_H^-|{\rm{in}}\rangle
\end{equation}
which, when written out explicitly, is the statement of the soft theorem, which is thus understood as the Ward identity of the corresponding symmetry. In~\cite{Freidel:2021dfs,Freidel:2021ytz,Freidel:2023gue} the authors write these charges for pure Yang-Mills and pure gravity in terms of creation and annihilation operators and explicitly compute the commutators. They were able to show that at leading and subleading order, the Ward identity takes the form
\begin{equation}
\langle{\rm{out}}|\mathcal{Q}^+_{H,i}\mathcal{S}-\mathcal{S}\mathcal{Q}_{H,i}^-|{\rm{in}}\rangle = \mathbb{D}_iS^{(i)}\langle\rm{out}|\mathcal{S}|\rm{in}\rangle
\end{equation}
where $S^{(i)}$ is the $i^{\rm{th}}$ order soft factor, to be explained below, and $\mathbb{D}_i$ is a differential operator that we leave unspecified here and will define in terms of specified coordinates later (see section \ref{sec:Charges}). 

In this paper we concentrate on the results in~\cite{Freidel:2021dfs}, where, from the perspective of canonically constructed charges, the authors aimed to understand the asymptotic symmetry associated with the sub-subleading soft graviton theorem, also referred to as the Cachazo-Strominger soft theorem~\cite{Cachazo:2014fwa}. Denoting an $n$-point graviton amplitude as $\mathcal{M}_n$, the sub-subleading soft theorem is given by 
\begin{equation}
\mathcal{M}_{n+1}^{(2)} =-\frac{1}{2}\sum_{a=1}^n\frac{\varepsilon_{\mu\nu}(q_\rho J_a^{\rho\mu})(q_\sigma J_a^{\sigma\nu})}{q\cdot k_a}\mathcal{M}_n = \frac{1}{2}\sum_{a=1}^n\frac{[sa]}{\langle sa\rangle}\tilde{\lambda}_s^2\partial_{\tilde{\lambda}_a}^2\mathcal{M}_n.
\end{equation}
In the first equality $\varepsilon_{\mu\nu}$ is the polarization tensor for the graviton, $q_\mu$ represents the momentum of the soft graviton, $k_a$ the is momentum of the hard particles and $J_a$ is the total angular momentum of the $a^{\rm{th}}$ particle. The second equality translates this formula into spinor-helicity notation which we will define and use throughout this paper. We care about this particular order of the soft graviton theorem because the separation of the charge into hard and soft parts has a caveat that we wish to understand from the point of view of scattering amplitudes. 

At sub-subleading order, in addition to the usual hard and soft charges, there is an additional component that the authors of \cite{Freidel:2021dfs} refer to as the collinear part of the charge. Although, they do not compute the commutator of this charge with the $\mathcal{S}$-matrix explicitly, they give the general form  
\begin{equation}\label{eq:collcharge}
\langle{\rm{out}}|\mathcal{Q}^+_C\mathcal{S}-\mathcal{S}\mathcal{Q}_C^-|{\rm{in}}\rangle \propto \sum_{a,b=1; a\neq b}^n F_{a b}(\omega_a,\omega_b) \delta^{(2)}(s,a)\delta^{(2)}(a,b)\lim_{a||b}\langle\rm{out}|\mathcal{S}|\rm{in}\rangle
\end{equation}
where $F_{ab}$ denotes an unspecified function of the energies of the two particles being summed over~($a,b$). The delta functions are in terms of the angular directions of the momenta of the soft particle  and the two particles being summed over. The right hand side of this equation contains a sum over the collinear limits of the $\mathcal{S}$-matrix and one would expect that these terms can be reproduced in the amplitudes context. Namely, if one finds collinear contributions to the sub-subleading soft graviton factor and then acts on them with the differential operator $\mathbb{D}_2$, it should reproduce the form on the right hand side above. 

In this paper, we reproduce these terms for all tree-level graviton amplitudes by looking at distributional terms in the soft expansion. In particular, derivatives in the soft factor acting on the amplitude produce distributional collinear terms. Since we are expanding an amplitude which does not contain distributional terms, we need to subtract these off in the definition of the soft factor. By this logic, we show that the sub-subleading soft graviton theorem should be corrected to\footnote{
In this paper we are always looking at the positive helicity soft theorem, i.e the soft particle has positive helicity. Of course, corresponding results exist in the opposite sector.
}
\begin{equation}
\mathcal{M}_{n+1}^{(2)}  = \frac{1}{2}\sum_{a=1}^n\frac{[sa]}{\langle sa\rangle}\tilde{\lambda}_s^2\partial_{\tilde{\lambda}_a}^2\mathcal{M}_n - \frac{1}{2}\sum_{a=1}^n\sum_{b>2;b>a}^n\frac{[sb]^3}{\langle sb\rangle}f_{h_P}^{h_ah_b}(t)\delta(a,b)\mathcal{M}_{n-1}(\ldots, P^{h_P},\ldots)
\end{equation}
where the first term is  the usual soft theorem and the second term subtracts off distributional collinear terms. Here $f_{h_P}^{h_ah_b}(t)$ is given by the graviton splitting functions, and $t$ is the fraction $\omega_a/(\omega_a+\omega_b)$ of the collinear energies.

In section~\ref{sec:prelim} we review the necessary aspects of the spinor helicity formalism, MHV amplitudes, soft theorems and collinear limits. In section~\ref{sec:spinor}, we identify the origin of distributional contributions as a simple, familiar derivative identity applied to spinor helicity variables. In section~\ref{sec:example} we show that for a low point example, the distributional terms at subleading order vanish, as expected from the asymptotic symmetry analysis\footnote{There are no distributional corrections at leading level, as the soft factor does not contain any derivatives.}. We also compute the distributional terms at sub-subleading level and demonstrate that they are consistent with the expectations from~\cite{Freidel:2021dfs}. Then, in section~\ref{sec:general}, we repeat the process for a general MHV graviton amplitude, arguing that the distributional terms vanish at the subleading level and are consistent with collinear charge terms at the sub-subleading level. In section \ref{sec:general tree level grav}, we argue that the result extends to general tree level amplitudes.  We conclude with a brief discussion of our results and future directions in section~\ref{sec:discussion}.

\section{Background and Conventions}\label{sec:prelim}
In this section, we outline the necessary background and conventions that we will use throughout this paper.
\paragraph{Spinor Helicity Formalism}
It will be convenient to use the spinor helicity formalism to write amplitudes in a clean form. In spinor helicity notation, momenta are written in terms of complex bispinors as 
\begin{equation}
    k_a^\mu \to k_a^{\alpha \dot\alpha}=\lambda_a^\alpha \tilde\lambda_a^{\dot\alpha},
\end{equation}
where the spinors $\lambda,\tilde\lambda$ are in the fundametal and anti-fundamental representation of $SL(2,\mathrm{C})$ respectively. For real momenta, the fundamental and anti-fundamental spinors are complex conjugates of one another. Spinor indices are raised and lowered with the anti-symmetric tensor, and spinor contractions are denoted by
\begin{equation}
[sa] \equiv \tilde{\lambda}_{s,\dot{\alpha}}\tilde{\lambda}_a^{\dot{\alpha}}, \ \ \langle sa\rangle \equiv\lambda_{s,\alpha}\lambda_a^{\alpha},
\end{equation}
and products of four-momenta take the form 
\begin{equation}
    2\,k_a\cdot k_b = \langle a b\rangle[ab ].
\end{equation}
The collinear limit corresponds to setting $(\lambda_a,\tilde\lambda_a)\propto (\lambda_b,\tilde\lambda_b)$, implying $k_a\cdot k_b =0 $. Following the notation of \cite{Cachazo:2014fwa}, we will let
\begin{equation}
    q^{\alpha\dot\alpha}=\lambda_s^\alpha \tilde\lambda_s^{\dot\alpha},
\end{equation}
denote the external graviton to be taken soft.

\paragraph{Soft Expansion} 

As mentioned above, graviton tree-level amplitudes have a well known divergent structure when an external graviton is taken soft. The leading term in the soft expansion goes back to Weinberg~\cite{Weinberg:1965nx} and is universal, including at loop level~\cite{Sen:2017nim}, while the sub- and sub-subleading level in the soft expansion were derived for tree-level graviton ampltiudes by Cachazo and Strominger~\cite{Cachazo:2014fwa}. Here, we briefly review the Cachazo-Strominger soft theorem.\footnote{For concreteness, we have chosen the outgoing graviton to have positive helicity. In the negative helicity case, one would replace $\lambda$ with $\tilde\lambda$ in the steps below.} Using the above spinor conventions, we denote a full amplitude by
\begin{equation}
\mathcal{M}_{n+1} = M_{n+1}\left(\{\lambda_1,\tilde{\lambda}_1,h_1\}, \cdots,\{\lambda_n,\tilde{\lambda}_n,h_n\}, \{\lambda_s,\tilde{\lambda}_s,h_s\}\right)\delta^{(4)}\left(\sum_{a=1}^n\lambda_{a,\alpha}\tilde{\lambda}_{a,\dot{\alpha}}+\lambda_{s,\alpha}\tilde{\lambda}_{s,\dot{\alpha}}\right),
\end{equation}
where $M_{n+1}$ is the stripped amplitude not containing the  momentum conserving delta function and $h_i$ denotes the helicity of an external particle. To take the soft limit one introduces a small expansion parameter $\epsilon$ that multiplies the momentum of the soft graviton
\begin{equation}
q_{\alpha,\dot{\alpha}}\rightarrow \epsilon q_{\alpha,\dot{\alpha}}=\epsilon\lambda_{s,\alpha}\tilde{\lambda}_{s,\dot{\alpha}}.
\end{equation}
The soft limit can be implemented as the following limit of the spinors, called the holomorphic soft limit,
\begin{equation}
\lambda_s\rightarrow \epsilon\lambda_s, \ \ \tilde{\lambda}_s\rightarrow \tilde{\lambda}_s,
\end{equation}
which is related to the more natural limit $\lambda_s\to \sqrt{\epsilon}\lambda_s,\,\tilde\lambda_s\to \sqrt{\epsilon} \tilde\lambda_s$ by an overall power of $\epsilon$ multiplying the amplitude. Using a BCFW-type expansion, the stripped amplitude can be expressed as a sum over lower-point amplitudes
\begin{equation}\label{eq:BCFW}
M_{n+1} = \frac{1}{\epsilon^3}\sum_{a=1}^{n-1}\frac{[sa]\langle na\rangle^2}{\langle sa\rangle\langle ns\rangle^2}M_n\left(\{\lambda_1,\tilde{\lambda}_1\}, \cdots,\{\lambda_a,\tilde{\lambda}_a + \epsilon\frac{\langle ns\rangle}{\langle na\rangle}\tilde{\lambda}_s\}, \cdots,\{\lambda_n,\tilde{\lambda}_n+\epsilon\frac{\langle as\rangle}{\langle an\rangle}\tilde{\lambda}_s\}\right)+\dots
\end{equation}
where $\lambda_n$ is a reference spinor chosen arbitrarily among the outgoing hard particles. The ellipsis indicate terms that are regular in the soft limit, which we omit here. One then expands the full amplitude around $\epsilon\to0$, giving
\begin{eqnarray}\label{eq:softexpansion1}
\mathcal{M}_{n+1}(\cdots,\{\epsilon\lambda_s,\tilde{\lambda}_s\}) & = & \frac{1}{\epsilon^3}\sum_{a=1}^{n-1}\frac{[sa]\langle na\rangle^2}{\langle sa\rangle\langle ns\rangle^2}\mathcal{M}_n\left(\{\lambda_i,\tilde{\lambda}_i\}\right)\cr
& + & \frac{1}{\epsilon^2}\sum_{a=1}^{n-1}\frac{[sa]\langle na\rangle^2}{\langle sa\rangle\langle ns\rangle^2}\left(\frac{\langle ns\rangle}{\langle na\rangle}\tilde{\lambda}_s\partial_{\tilde{\lambda}_a}+\frac{\langle as\rangle}{\langle an\rangle}\tilde{\lambda}_s\partial_{\tilde{\lambda}_n}\right)\mathcal{M}_n\left(\{\lambda_i,\tilde{\lambda}_i\}\right)\cr
& + & \frac{1}{\epsilon}\sum_{a=1}^{n-1}\frac{[sa]\langle na\rangle^2}{2\langle sa\rangle\langle ns\rangle^2}\left(\frac{\langle ns\rangle}{\langle na\rangle}\tilde{\lambda}_s\partial_{\tilde{\lambda}_a}+\frac{\langle as\rangle}{\langle an\rangle}\tilde{\lambda}_s\partial_{\tilde{\lambda}_n}\right)^2\mathcal{M}_n\left(\{\lambda_i,\tilde{\lambda}_i\}\right)\cr
& + & \mathcal{O}(1),
\end{eqnarray}
where $\tilde\lambda_i \partial_{\tilde\lambda_j}\equiv\tilde\lambda_i^{\dot\alpha}\frac{\partial}{\partial\tilde\lambda_j^{\dot\alpha}}$.
Using momentum conservation and conservation of angular momentum, the expansion can be expressed in a a more compact form\footnote{Going from \eqref{eq:softexpansion1} to \eqref{eq:softexpansion2} uses $\partial_{\tilde\lambda_a}\langle a b \rangle^{-1} = 0$, while we later on treat these derivatives as non-vanishing distributional terms. This should not cause any confusion. One assumes this derivative vanishes in~\eqref{eq:softexpansion1} to have compact expression. However, going from \eqref{eq:BCFW} to \eqref{eq:softexpansion2} does not at all rely on this assumption.
}
\begin{eqnarray}\label{eq:softexpansion2}
\mathcal{M}_{n+1}(\cdots,\{\epsilon\lambda_s,\tilde{\lambda}_s\}) & = & \frac{1}{\epsilon^3}\sum_{a=1}^{n-1}\frac{[sa]\langle na\rangle^2}{\langle sa\rangle\langle ns\rangle^2}\mathcal{M}_n\left(\{\lambda_i,\tilde{\lambda}_i\}\right)\cr
& + & \frac{1}{\epsilon^2}\sum_{a=1}^{n-1}\frac{[sa]\langle na\rangle}{\langle sa\rangle\langle ns\rangle}\tilde{\lambda}_s\partial_{\tilde{\lambda}_a}\mathcal{M}_n.\left(\{\lambda_i,\tilde{\lambda}_i\}\right)\cr
& + & \frac{1}{\epsilon} 
\sum_{a=1}^{n}\frac{[sa]}{2\langle sa\rangle} \tilde{\lambda}_s^2\partial_{\tilde{\lambda}_a}^2\mathcal{M}_n\left(\{\lambda_i,\tilde{\lambda}_i\}\right)\cr
& + & \mathcal{O}(1),
\end{eqnarray}
where the prefactor in each term is expressed in terms of external momenta, the polarization tensor of the soft graviton and the angular momentum operator when translating back from spinor helicity notation.\footnote{Note that the differential operator only acts on the stripped part of the amplitude. The delta functions on both sides of the equation includes $n+1$ particle momenta, one of which is proportional to $\epsilon$. }

\paragraph{Collinear limits} 

Following the notation of \cite{Dixon:2013uaa}, we perform the collinear limit $k_{a} || k_{b}$ by taking the spinors parallel in the following way
\begin{equation}
(\lambda_a,\tilde{\lambda}_a)\rightarrow \sqrt{t}(\lambda_P,\tilde{\lambda}_P), \ \ (\lambda_b,\tilde{\lambda}_b)\rightarrow \sqrt{1-t}(\lambda_P,\tilde{\lambda}_P),
\end{equation}
such that $k_a=t\, k_P$ and $k_b=(1-t) k_P$, and $k_P^2 = 2 k_a\cdot k_b =0$. In the collinear limit, graviton amplitudes factorize into a splitting function and a lower point amplitude, \textit{viz.} \cite{Bern:1998sv} 
\begin{equation}
    M_n(\dots,a^{h_a},b^{h_b},\dots) \overset{a||b}{\longrightarrow} \sum_{h}\text{Split}_{h}^{\text{grav}}(t,a^{h_a},b^{h_b})\times M_{n-1}(\dots, P^h,\dots),
\end{equation}
%\enote{Note to self: is it ok using $h$ as a label here?}
where the graviton splitting functions take the form 
\begin{eqnarray}\label{eq:splitting}
\mbox{Split}_-(t,a^+,b^+) = 0, \ \ \mbox{Split}_+(t,a^+,b^+) = -\frac{1}{t(1-t)}\frac{[ab]}{\langle ab\rangle}, \ \ \mbox{Split}_-(t,a^-,b^+) = -\frac{t^3}{1-t}\frac{[ab]}{\langle ab\rangle}.
\end{eqnarray}
Note that the splitting functions in gravity are not divergent in the the collinear limit\footnote{A significant portion of the recent literature chooses to use $(2,2)$ signature where the spinors are real and independent. If one does this, then the splitting functions can be divergent since we can take $\langle ab\rangle\rightarrow 0$ while keeping $[ab]$ fixed. This provides a better interpretation of these splitting functions as OPE coefficients in the boundary theory, as discussed in~\cite{Pate:2019lpp}.} since $\langle a b\rangle^*=[a b]$. In what follows, it will be notationally convenient to write $\mbox{Split}_{h_P}(t,a^{h_a},b^{h_b}) = f_{h_P}^{h_ah_b}(t)\frac{[ab]}{\langle ab\rangle}$ where $h_P,h_a,h_b$ are the helicities of the particles.

\paragraph{MHV Amplitudes}
It will be helpful to work with explicit expressions for graviton amplitudes. For this purpose, we will consider MHV amplitudes. These are amplitudes with two negative helicity  and $n-2$ positive helicity external gravitons. MHV amplitudes can be expressed in a compact form provided by Hodges' formula~\cite{Hodges:2012ym} as
\begin{equation}
M_n  =  \frac{\langle 12\rangle^6}{\langle 23\rangle^2\langle 31\rangle^2}\mbox{det}\Phi_{123}^{123} = \frac{\langle 12\rangle^6}{\langle 23\rangle^2\langle 31\rangle^2}\sum_{\{i_4,\cdots,i_n\}\in\{4,\cdots n\}}\varepsilon_{i_4\cdots i_n}\Phi_{4,i_4}\cdots\Phi_{n,i_n}
\end{equation}
where 
\begin{equation}
\Phi_{ab} = \frac{[ab]}{\langle ab\rangle}, \ a\neq b, \ \ \Phi_{ab} = -\sum_{c=1;c\neq a}^n\frac{[ac]}{\langle ac\rangle}\frac{\langle 1c\rangle\langle 2c\rangle}{\langle 1a\rangle\langle 2a\rangle}, \ a=b.
\end{equation}
As we show in appendix \ref{app:equiv}, these amplitudes can also be expressed on the from 
\begin{equation}
    M_n = \mathcal{F}_n\left(\prod_{i=1, j>i}^n\frac{1}{\langle ij\rangle}\right),
\end{equation}
where the factor $\mathcal{F}_n$ does not contain any poles. This form will be convenient when we consider distributional terms arising from derivatives of poles in the amplitude. 

\section{Distributional Terms from Spinor Derivatives}\label{sec:spinor}

As we have seen, the subleading and sub-subleading soft theorems require us to take derivatives of the amplitude with respect to the spinor variables. In particular, for the positive helicity soft theorem we get expressions involving the differential operator $\tilde\lambda_i \partial_{\tilde\lambda_j}$. Acting on square brackets gives
\begin{equation}
\tilde{\lambda}_i\partial_{\tilde{\lambda}_j}[jn] = \tilde{\lambda}_i\partial_{\tilde{\lambda}_j}(\tilde{\lambda}_j\tilde{\lambda}_n) = \tilde{\lambda}_i\tilde{\lambda}_n = [in],
\end{equation}
while it vanishes acting on angle brackets. However, since the components of $\lambda$ and $\tilde\lambda$ are conjugates of each other, we have to be mindful of the following identity, see e.g.~\cite{DiFrancesco:639405},
\begin{equation}
   \frac{\partial}{\partial \tilde\lambda ^\alpha}\frac{1}{\lambda^\alpha} =\pi \delta^2(\lambda^\alpha),
\end{equation}
where the delta function is defined by $\delta^2(\lambda^\alpha)=\delta(\text{Re}\lambda^\alpha)\delta(\text{Im}\lambda^\alpha)$. This identity follows from a straightforward application of Stokes' theorem and Cauchy's theorem. This gives an identity for $\tilde{\lambda}_i\partial_{\tilde{\lambda}_j}$ acting on $\langle ab \rangle^{-1}$, namely
\begin{equation}
\begin{split}
    \tilde\lambda_i \partial_{\tilde\lambda_j} \frac{1}{\langle j k\rangle}&=\left(\tilde\lambda_i ^0\frac{\partial}{\partial \tilde\lambda^0_j}+\tilde\lambda_i ^1\frac{\partial}{\partial \tilde\lambda^1_j}\right)\frac{1}{\lambda^0_j \lambda^1_k -\lambda^1_j\lambda^0_k }\\[10pt]
    &= \pi [i k] \,\delta^2(\langle jk \rangle).
\end{split}
\end{equation}
Note that $\delta^2(\langle ab \rangle)$ enforces $\langle ab \rangle=[ab]=0$ since $\langle ab \rangle=[ab]^*$. To make the notation more symmetric between square and angle brackets, and to not carry factors of $\pi$ around, we will use the notation
\begin{equation}
    \delta(j,k)\equiv\pi \delta^2(\langle j k \rangle).
\end{equation}
so we have 
\begin{equation}
       \tilde\lambda_i \partial_{\tilde\lambda_j} \frac{1}{\langle j k\rangle} = [i k]\,\delta(j,k)
\end{equation}
These distributional terms are usually dropped~\cite{Cachazo:2014fwa} when taking spinor derivatives. However, in principle, they should be subtracted from the soft factor such that both sides of~\eqref{eq:softexpansion2} are without distributional terms. Below, we will find these distributional terms for the Cachazo-Strominger soft theorem and find that they correspond to the collinear charge component found in~\cite{Freidel:2021ytz}.

\section{Example: MHV$_5$}\label{sec:example} 

In what follows we will discuss, in detail, a low point example to illustrate the existence and form of the distributional terms. In particular we show that the leading and subleading soft factors do not produce any such terms, while the sub-subleading case does. We will consider the five-point MHV graviton amplitude, so we need the explicit form of four- and  five-point amplitudes. They are given in the spinor-helicity variables as 
\begin{equation}
\mathcal{M}_4^{--++} = \frac{\langle 12\rangle^6[43]}{\langle 23\rangle\langle 13\rangle\langle 43\rangle\langle 14\rangle\langle 24\rangle}, \ \ \mathcal{M}_5^{--+++} = \frac{\langle 12\rangle^7}{\langle 13\rangle\langle 23\rangle\langle 34\rangle\langle 35\rangle\langle 45\rangle}\left[\frac{[14]}{\langle 14\rangle}\frac{[25]}{\langle 25\rangle} - \frac{[15]}{\langle 15\rangle}\frac{[24]}{\langle 24\rangle}\right].
\end{equation}
Once we have found the form of  the distributional terms at sub-subleading level for this example, we will move on to the general case.~\footnote{The calculation of all the distributional terms is quite tedious. We used Mathematica to help simplify them and have included some details in the appendices.}

Note that we need to work with the $n$-point amplitude of degree $n\geq 4$ to have non-trivial distributional terms. To see this, suppose that the soft factor was acting on the three-point amplitude. Then we have particles $1,2,3,s$, and there will be distributional terms containing $\delta(a,b)$ for $a,b\in \{1,2,3\}$. In front of this term there will be a factor of $\langle 12\rangle$. If $a,b=1,2$ the delta function will vansih, so the only cases we need to look at are the cases where $a=1,2$ and $b=3$. 
\begin{eqnarray}
\langle 12\rangle \delta(1,3) & = & \frac{\langle 12\rangle[2s]}{[2s]}\delta(1,3) = -\frac{\langle 13\rangle[3s]}{[2s]}\delta(1,3) = 0\cr
\langle 12\rangle \delta(2,3) & = & \frac{\langle 12\rangle[1s]}{[1s]}\delta(2,3) = \frac{\langle 23\rangle[3s]}{[1s]}\delta(2,3) = 0.
\end{eqnarray}
We see that in this case, we can always make the distributional terms go away using momentum conservation. This does not work at higher orders, which is why the four-point amplitude is not a suitable example and we take the five-point amplitude as our starting point. 

\subsection{Leading Soft Factor}
The leading soft theorem is given by 
\begin{equation}
\mathcal{M}_{n+1}^{(0)} = \sum_{a=1}^{n-1}\frac{[sa]\langle na\rangle^2}{\langle sa\rangle\langle ns\rangle^2}\mathcal{M}_n
\end{equation}
where $s$ labels the soft momentum and the superscript $(0)$ denotes that it is the leading level expression. There are no derivatives acting on the amplitude here so there are no new distributional terms. Therefore we simply get the usual result
\begin{equation}
\mathcal{M}_5^{(0)} = \left[\frac{[51]\langle 14\rangle^2}{\langle 51\rangle\langle 54\rangle^2} + \frac{[52]\langle 24\rangle^2}{\langle 52\rangle\langle 54\rangle^2}+ \frac{[53]\langle 34\rangle^2}{\langle 53\rangle\langle 54\rangle^2}\right]\mathcal{M}_4.
\end{equation}
\subsection{Subleading Soft Factor}
The subleading soft theorem is given by 
\begin{equation}
\mathcal{M}_{n+1}^{(1)} = \sum_{a=1}^{n-1}\frac{[sa]\langle na\rangle}{\langle sa\rangle\langle ns\rangle}\tilde{\lambda}_s\partial_{\tilde{\lambda}_a}\mathcal{M}_n.
\end{equation}
In this case, since we have derivatives acting on the amplitude, we will have distributional terms. However, as shown in Appendix \ref{sec: Cancel_subleading_MHV_4}, one can use momentum conservation to put all distributional terms on the form $z\delta(z)$, which vanishes, giving 
\begin{equation}
\mathcal{M}_5^{(1)} = \frac{[45][53]\langle 43\rangle}{[43]\langle 45\rangle \langle 53\rangle}\mathcal{M}_4.
\end{equation}
Letting $\epsilon=1$ and summing the leading and subleading contributions can be shown to reproduce the exact five point amplitude, as expected. We show this in Appendix \ref{sec: full_five_point} to confirm consistency with \cite{Cachazo:2014fwa}.

\subsection{Distributional Terms for the Sub-subleading Soft Factor}\label{sec:distributional}
The sub-subleading soft theorem is a different story because the distributional terms do not vanish. As we reviewed in section \ref{sec:prelim}, the sub-subleading soft factor can be written in spinor helicity terms as
\begin{equation}
\mathcal{M}_{n+1}^{(2)}  = \frac{1}{2}\sum_{a=1}^n\frac{[sa]}{\langle sa\rangle} \mathcal{D}_a^2\mathcal{M}_n,
\end{equation} where from here on forward, we will write $\mathcal{D}_a \equiv \tilde{\lambda}_s\partial_{\tilde{\lambda}_a} $ to simplify the notation.
In addition to the usual soft factor, there will be distributional terms arising from the $\mathcal{D}_a^2$ acting on poles in the amplitude. For the example we are considering, the non-distributional terms vanish, as expected~\cite{Cachazo:2014fwa}. In principle, there will be distributional terms involving delta functions, products of delta functions and derivatives of delta functions. We compute these terms explicitly, then simplify them in Appendix \ref{sec: simplify_dist_MHV_4} to obtain
\begin{eqnarray}
\mathcal{M}_5^{(2)} & = & -\frac{\langle 12\rangle^6[15][35][45]}{2\langle 14\rangle\langle 23\rangle\langle 24\rangle\langle 34\rangle\langle 35\rangle}\delta(1,3) -  \frac{\langle 12\rangle^6[25][35] [45]}{2\langle 13\rangle\langle 14\rangle\langle 24\rangle\langle 34\rangle\langle 35\rangle}\delta(2,3) \cr
& + & \frac{\langle 12\rangle^6[15][45][35]}{2\langle 13\rangle\langle 23\rangle\langle 24\rangle\langle 34\rangle\langle 45\rangle}\delta(1,4) +  \frac{\langle 12\rangle^6[25][45][35]}{2\langle 13\rangle\langle 14\rangle\langle 23\rangle\langle 34\rangle\langle 45\rangle}\delta(2,4) \cr
& + & \frac{\langle 12\rangle^6[35][45](\langle 35\rangle[35]+\langle 45\rangle[45])}{2\langle 13\rangle\langle 14\rangle\langle 23\rangle\langle 24\rangle\langle 35\rangle\langle 45\rangle}\delta(3,4),
\end{eqnarray}
where we see that we end up with terms involving only a single delta function. 
This form is not particularly enlightening, until  one imposes the delta function $\delta(a,b)$, which makes the particles $a,b$ collinear. As in section \ref{sec:prelim}, we take $a,b$ collinear by taking their spinors to 
\begin{equation}\label{eq:spincoll}
(\lambda_a,\tilde{\lambda}_a)\rightarrow \sqrt{t}(\lambda_P,\tilde{\lambda}_P), \ \ (\lambda_b,\tilde{\lambda}_b)\rightarrow \sqrt{1-t}(\lambda_P,\tilde{\lambda}_P),
\end{equation}
which gives the result
\begin{eqnarray}
\mathcal{M}_5^{(2)} & = & \frac{t^3}{2(1-t)}\frac{[P5]^3}{\langle P5\rangle}\frac{\langle P2\rangle^6}{\langle 24\rangle^2\langle P4\rangle^2}\delta(1,3) +  \frac{t^3}{2(1-t)}\frac{[P5]^3}{\langle P5\rangle}\frac{\langle 1P\rangle^6}{\langle 14\rangle^2\langle P4\rangle^2}\delta(2,3) \cr
& + & \frac{t^3}{2(1-t)}\frac{[P5]^3}{\langle P5\rangle}\frac{\langle P2\rangle^6}{\langle 23\rangle^2\langle P3\rangle^2}\delta(1,4)  +  \frac{t^3}{2(1-t)}\frac{[P5]^3}{\langle P5\rangle}\frac{\langle 1P\rangle^6}{\langle 13\rangle^2\langle P3\rangle^2}\delta(2,4) \cr
& + & \frac{1}{2t(1-t)}\frac{[P5]^3}{\langle P5\rangle}\frac{\langle 12\rangle^6}{\langle 1P\rangle^2\langle 2P\rangle^2}\delta(3,4). 
\end{eqnarray}
We can write this more compactly by noting that \begin{equation}
\mathcal{D}_a^2\left[\frac{[ab]}{\langle ab\rangle}\right] = \mathcal{D}_a\left[\frac{[sb]}{\langle ab\rangle} + [ab][sb]\delta(a,b)\right] = [sb]^2\delta(a,b), 
\end{equation}
and also that $\frac{[P5]}{\langle P5\rangle} = \frac{[a5]}{\langle a5\rangle} = \frac{[b5]}{\langle b5\rangle}$, which allows us to write 
\begin{equation}\label{eq:derivofsplit}
\mathcal{M}_5^{(2)} = \frac{1}{2}\sum_{a=1}^n\sum_{b>2;b>a}^n\frac{[5b]}{\langle 5b\rangle}\mathcal{D}_a^2\left(\mbox{Split}_{h_P}(t,a^{h_a},b^{h_b})\right)\mathcal{M}_{3}(\ldots P^{h_P}\ldots).
\end{equation}
We can see that for this low point case, the distributional terms are the same as the ones we would obtain by first taking collinear limits of the amplitude and then applying the soft factor.

\section{Proof for General MHV Amplitude}\label{sec:general}
While it was instructive to see an explicit example, we would like to show that these distributional terms arise in the case of a general graviton amplitude. Soft and collinear limits involve expanding an amplitude order by order, either in soft momentum or collinear momenta. When we discuss these limits in the context of amplitudes, we are often concentrating on a particular order in these expansions, namely leading order in the collinear case. It is thereby natural to question whether the order of soft and collinear limits matters for a given scattering amplitude. 

In light of this, we will first consider a general MHV graviton amplitude, to confirm that there is no ambiguity in taking the soft and collinear limits, as was already the case in the example above. Then we will argue that the distributional terms take the same form for a general graviton amplitude. We will use the form of the MHV graviton amplitude that we have shown is equivalent to Hodges' formula
\begin{equation}
M_n = \mathcal{F}_n\left(\prod_{i=1, j>i}^n\frac{1}{\langle ij\rangle}\right)
\end{equation}
where the function $\mathcal{F}_n$ has no poles and is essentially the Hodges determinant written over a common denominator. We will need the action of the spinor derivative on this form. In particular we will need
\begin{equation}
\mathcal{D}_a\left(\prod_{i=1, j>i}^n\frac{1}{\langle ij\rangle}\right) = \sum_{k=1,\neq a}^n\langle ak\rangle [sk]\delta(a,k)\left(\prod_{i=1, j>i}^n\frac{1}{\langle ij\rangle}\right)
\end{equation}
where the $\langle ak \rangle$ cancels out the pole that gave rise to the delta function. We write it this way to preserve the form of the product of poles because it makes it easier to keep track of the cancellations. It will also be useful to have the second derivative
\begin{eqnarray}
\mathcal{D}_a^2\left(\prod_{i=1, j>i}^n\frac{1}{\langle ij\rangle}\right) & = & \sum_{k=1,\neq a}^n\langle ak\rangle [sk]\mathcal{D}_a \delta(a,k)\left(\prod_{i=1, j>i}^n\frac{1}{\langle ij\rangle}\right) \cr
& + & \sum_{k,\ell=1,\neq a}^n\langle ak\rangle\langle a\ell\rangle [sk][s\ell]\delta(a,k) \delta(a,\ell)\left(\prod_{i=1, j>i}^n\frac{1}{\langle ij\rangle}\right).
\end{eqnarray}
First we will show that the distributional terms vanish in the subleading case.

\subsection{Subleading Soft}

In the case of the subleading soft theorem, we have
\begin{eqnarray}
\mathcal{M}_{n+1}^{(1)} & = & \sum_{a=1}^{n-1}\frac{[sa]\langle na\rangle}{\langle sa\rangle\langle ns\rangle}\mathcal{D}_a\left(\mathcal{F}_n\left(\prod_{i=1, j>i}^n\frac{1}{\langle ij\rangle}\right)\right) \cr
& = & \sum_{a=1}^{n-1}\frac{[sa]\langle na\rangle}{\langle sa\rangle\langle ns\rangle}\left(\mathcal{D}_a\mathcal{F}_n\right)\left(\prod_{i=1, j>i}^n\frac{1}{\langle ij\rangle}\right)+ \sum_{a=1}^{n-1}\frac{[sa]\langle na\rangle}{\langle sa\rangle\langle ns\rangle}\mathcal{F}_n\mathcal{D}_a\left(\prod_{i=1, j>i}^n\frac{1}{\langle ij\rangle}\right).
\end{eqnarray}
The first term is the usual non-distributional contribution to the subleading soft theorem. We only need to look at the second term in more detail which expands to 
\begin{eqnarray}
\mathcal{M}_{n+1,\rm{dist}}^{(1)} & = & \sum_{a=1}^{n-1}\frac{[sa]\langle na\rangle}{\langle sa\rangle\langle ns\rangle}\mathcal{F}_n\sum_{k=1,\neq a}^n\langle ak\rangle[sk]\delta(a,k)\left(\prod_{i=1, j>i}^n\frac{1}{\langle ij\rangle}\right).
\end{eqnarray}
The distributional terms will come in pairs that share the same delta function and the coefficient of each independent delta function has to vanish independently. Looking at just one pair
\begin{equation}
\mathcal{M}_{n+1,{\rm{dist}},ak}^{(1)}  = \mathcal{F}_n\frac{\langle ak\rangle[sa][sk]}{\langle ns\rangle}\left[\frac{\langle na\rangle}{\langle sa\rangle} - \frac{\langle nk\rangle}{\langle sk\rangle}\right]\delta(a,k)\left(\prod_{i=1, j>i}^n\frac{1}{\langle ij\rangle}\right),
\end{equation}
we see that the term in square brackets vanishes by the Schouten identity. Therefore all distributional terms are 0 in the subleading case.

We can alternatively write this as 
\begin{eqnarray}
\mathcal{M}_{n+1,\rm{dist}}^{(1)} & = & \sum_{a=1}^{n-1}\frac{[sa]\langle na\rangle}{\langle sa\rangle\langle ns\rangle}\sum_{k=1,\neq a}^n\langle ak\rangle[sk]\delta(a,k)\mathcal{M}_n.
\end{eqnarray}
 Noting that in the collinear limit $\mathcal{M}_n\rightarrow \frac{[ak]}{\langle ak\rangle}\mathcal{M}_{n-1}$, we see that this vanishes without using the Schouten identity argument above, since each term contains $[ak]\delta(a,k)$. Therefore, we see that there are two independent ways to show that the subleading distributional terms vanish. This is consistent with the four point example in the previous section.

\subsection{Sub-subleading Soft}

Now we need to show what happens in the case of the sub-subleading soft theorem. 
\begin{eqnarray}
\mathcal{M}_{n+1}^{(2)} & = & \frac{1}{2}\sum_{a=1}^n\frac{[sa]}{\langle sa\rangle}\mathcal{D}_a^2\mathcal{M}_n = \frac{1}{2}\sum_{a=1}^n\frac{[sa]}{\langle sa\rangle}\mathcal{D}_a^2\mathcal{F}_n\left(\prod_{i=1, j>i}^n\frac{1}{\langle ij\rangle}\right) \cr
& + &  \sum_{a=1}^n\frac{[sa]}{\langle sa\rangle}\mathcal{D}_a\mathcal{F}_n\sum_{k=1,\neq a}^n\langle ak\rangle [sk]\delta(a,k)\left(\prod_{i=1, j>i}^n\frac{1}{\langle ij\rangle}\right)\cr
& + &  \frac{1}{2}\sum_{a=1}^n\frac{[sa]}{\langle sa\rangle}\sum_{k=1,\neq a}^n\langle ak\rangle [sk]^2\partial \delta(a,k)\mathcal{M}_n\cr
& + & \frac{1}{2}\sum_{a=1}^n\frac{[sa]}{\langle sa\rangle}\sum_{k,\ell=1,\neq a}^n\langle ak\rangle\langle a\ell\rangle [sk][s\ell]\delta(a,k) \delta(a,\ell)\mathcal{M}_n
\end{eqnarray}
The first term is the usual non-distributional part of the sub-subleading soft theorem, as the derivatives act on a function that is completely regular. We turn our attention to the terms containing delta functions. As we saw in the four point example, there are three types of delta function terms. To simplify these terms, it is important to recall that in the collinear limit $a||b$, the graviton splitting function goes like $\frac{[ab]}{\langle ab\rangle}$. We see that the double delta function term will vanish because in the collinear limit it contains factors of $[ak]\delta(a,k)$ or $[a\ell]\delta(a,\ell)$. This is not surprising because these terms vanished in this way in the four point case. In the case of the derivative term, in the collinear limit, the angle bracket will cancel with the denominator of the splitting function, and we can simplify the result using $[ak]\partial \delta(a,k)\rightarrow -\delta(a,k)$, so these terms will all add to the single derivative term.

Taking the collinear limits and using the fact that $\mathcal{F}_n \to [ak] \mathcal{F}_{n-1}$ , we see that the distributional terms above recombine to give 
\begin{eqnarray}
\mathcal{M}_{n+1,\, distr.}^{(2)} & = & \frac{1}{2}\sum_{a=1}^n\frac{[sa]}{\langle sa\rangle}\sum_{k=1,\neq a}^n [sk]^2 \delta(a,k)f_{h_P}^{h_ah_k}(t)\mathcal{M}_{n-1}(\ldots, P^{h_P},\ldots)\cr
& + & \sum_{a=1}^n\frac{[sa]}{\langle sa\rangle}\mathcal{D}_a\mathcal{F}_{n-1}\sum_{k=1,\neq a}^n [ak]f_{h_P}^{h_ah_k}(t)[sk]\delta(a,k)\left(\langle ak\rangle\prod_{i=1, j>i}^n\frac{1}{\langle ij\rangle}\right) \cr
& = & \frac{1}{2}\sum_{a=1}^n\frac{[sa]}{\langle sa\rangle}\sum_{k=1,\neq a}^n [sk]^2 \delta(a,k)f_{h_P}^{h_ah_k}(t)\mathcal{M}_{n-1}(\ldots, P^{h_P},\ldots)\cr
& = & \frac{1}{2}\sum_{a=1}^n\sum_{k=1,\neq a}^n \frac{[sk]}{\langle sk\rangle}\mathcal{D}_a^2\left(\mbox{Split}_{h_P}(t,a^{h_a},k^{h_k})\right)\mathcal{M}_{n-1}(\ldots, P^{h_P},\ldots)
\end{eqnarray}
We see that in the general case, for a tree-level MHV amplitude we obtain the same form as we did in the low point example in~\eqref{eq:derivofsplit}. Therefore, for all tree-level MHV amplitudes the distributional terms in the soft theorem are consistent with first taking the collinear limit of the amplitude and then the soft limit.

\section{General Tree-level Graviton Amplitudes}\label{sec:general tree level grav}

Although it was helpful to work with a closed form expression in the case of MHV amplitudes for a fully detailed proof, one expects the result to carry over to general tree-level amplitudes\footnote{We thank Freddy Cachazo for pointing this out.}. The distributional terms are always supported in the collinear limit, where the amplitude factorizes, so such terms should, in general, arise from the soft factor acting on the splitting function in the factorized amplitude, as above. Therefore, when the soft factor can be written as a power of $\mathcal{D}_a$ acting on the lower point amplitude, we can subtract the distributional terms by shifting the derivative that appears in the soft factor
\begin{equation}
    (\mathcal{D}_a )^i\mathcal{M}_n \to  (\mathcal{D}_a )^i\mathcal{M}_n -   \sum_{k=1,\neq a}^n\mathcal{D}_a^i\left(\mbox{Split}_{h_P}(t,a^{h_a},k^{h_k})\right)_{\text{distr.}}\mathcal{M}_{n-1}.
\end{equation}
The subscript ``distr." indicates that we are just subtracting off distributional terms - e.g.~there are no terms present when $i=0$. In the case of soft gravitons, $i\in\{0,1,2\}$ because beyond sub-subleading order it is not clear that the soft factor can be written as a power of $\mathcal{D}_a$.\footnote{Beyond sub-subleading order in gravity, the soft theorems are not universal but rather have a universal part. It is likely that the universal part of those soft theorems can be written as a power of $\mathcal{D}_a$ acting on the lower point amplitude. It is therefore possible that this shift in $\mathcal{D}_a$ reproduces the collinear charges since they should correspond to the universal part of the soft theorem but we do not work this out explicitly here.} The calculations in the previous section show that these terms vanish at subleading level and give only single collinear delta function terms at sub-subleading level. In particular, for the sub-subleading soft theorem we get
\begin{equation}
\mathcal{M}_{n+1}^{(2)}  = \frac{1}{2}\sum_{a=1}^n\frac{[sa]}{\langle sa\rangle}\mathcal{D}_a^2\mathcal{M}_n - \frac{1}{2}\sum_{a=1}^n\sum_{b>2;b>a}^n\frac{[sb]^3}{\langle sb\rangle}f_{h_P}^{h_ah_b}(t)\delta(a,b)\mathcal{M}_{n-1}(\ldots, P^{h_P},\ldots),
\end{equation}
where both sides of the equation are now free of distributional terms. 

\section{Connection to Canonical Charges}\label{sec:Charges}

 In this section, we show that the collinear corrections to the sub-subleading soft factor reproduces the collinear terms found in the asymptotic symmetry analysis of~\cite{Freidel:2021dfs}. Doing so requires the introduction of some slightly different notation. As in~\cite{Freidel:2021dfs}, it is common to parametrize the momentum of massless particles in terms of an energy $\omega$ and two angular variables $z,\bar{z}$ as 
\begin{equation}
p^\mu = \omega(1+z\bar{z},z+\bar{z},-i(z-\bar{z}),1-z\bar{z}).
\end{equation}
Here $z,\bar{z}$ are complex coordinates that denote a position on the boundary two-dimensional sphere. In order to compare the soft factor to the action of the canonical charge in the sub-subleading case, one needs to first write the corrected soft theorem in terms of the sphere coordinates 
\begin{equation}
{S'}_+^{(2)}\langle \mbox{out}|\mathcal{S}|\mbox{in}\rangle  =  S_+^{(2)}\langle \mbox{out}|\mathcal{S}|\mbox{in}\rangle - \sum_{a=1}^n\sum_{b>2;b>a}^n\frac{\omega\omega_b\left(\bar{z}-\bar{z}_b\right)^3}{2(z-z_b)}f_{h_P}^{h_ah_b}(t)\delta(a,b)\langle(\omega_a+\omega_b)\hat{p}_a|\mathcal{S}|\mbox{in}\rangle
\end{equation}
and then take four derivatives with respect to the soft $\bar{z}$
\begin{equation}
\partial_{\bar{z}}^4{S'}_+^{(2)}\langle \mbox{out}|\mathcal{S}|\mbox{in}\rangle  =  \partial_{\bar{z}}^4S_+^{(2)}\langle \mbox{out}|\mathcal{S}|\mbox{in}\rangle - \frac{1}{2}\sum_{a=1}^n\sum_{b>2;b>a}^nf^{h_ah_b}_{h_P}(t)\delta(s,b)\delta(a,b)\langle(\omega_a+\omega_b)\hat{p}_a|\mathcal{S}|\mbox{in}\rangle.
\end{equation}
We can see that the first term is the usual sub-subleading hard charge and the second term is the non-divergent collinear contribution as found in~\cite{Freidel:2021dfs}.

\section{Discussion}\label{sec:discussion}

We have argued that the sub-subleading soft theorem for tree-level graviton amplitudes contains a distributional correction that is consistent with the form of the collinear component of the asymptotic charge found in~\cite{Freidel:2021dfs}. In order to avoid potential ambiguities with the order of soft and collinear limits, we have first shown rigorously for tree-level MHV amplitudes that taking the soft limit and then imposing the collinear constraints from the distributional terms is equivalent to acting with the soft factor on the amplitude around the corresponding pole. We then extend the argument to all tree-level graviton amplitudes, as they all have the same collinear splitting function. It would be interesting to see whether there are more complex distributional terms that appear at higher order terms in the soft expansion and what implications that has for the algebra of the corresponding charges. Our expectation is that the collinear contributions in the higher charges correspond to collinear limits of the universal part of the higher order soft theorems.

It should be noted that there is also a divergent contribution to the collinear charge in~\cite{Freidel:2021dfs} which contains two soft graviton creation operators. From an amplitudes context, we do not expect to see such a contribution since it is not a feature of the amplitude itself in the presence of a single soft limit.\footnote{We thank Laurent Freidel and Ana Raclariu for this observation.} Therefore, we do not attempt to resolve this in this paper but it would be interesting to understand that contribution in more detail.

It is also well known that in gravity, the soft theorems beyond leading order get corrected at loop level~\cite{Bern:2014oka} while the leading collinear splitting function remains uncorrected~\cite{Bern:1998sv}. It might be instrumental to consider self-dual gravity as a playground to study this since there are a finite set of amplitudes whose soft and collinear structure is well known~\cite{Bern:1998xc}. Since the soft factor has a different structure in terms of the derivative operators, it is likely that at loop order there will be additional distributional terms to what we have encountered at tree-level.

Finally, while our story is in gravity, there should be an analogous construction in gauge theory. In principle it should be easier to show since the MHV gluon amplitudes have a simple form. We expect that the collinear terms would appear at subleading level in gauge theory and also arise from the derivative acting on the splitting function. We hope to explore this in future work. 

\section*{Acknowledgements}
 The authors would like to thank Freddy Cachazo, Laurent Freidel, Yangrui Hu, Rajamani Narayanan, Sabrina Pasterski and Ana Raclariu for useful discussions. In particular, we would like to acknowledge Ana Raclariu for bringing this issue to our attention. S.N. acknowledges support by the Celestial Holography Initiative at the Perimeter Institute for Theoretical Physics and by the Simons Collaboration on Celestial Holography. E.J. is supported in part by funding from the Natural Sciences and Engineering Research Council of Canada, and from the BMO Financial Group. The authors' research at the Perimeter Institute is supported by the Government of Canada through the Department of Innovation, Science and Industry Canada and by the Province of Ontario through the Ministry of Colleges and Universities.

\appendix
\section{Cancellation of distributional terms: Subleading on MHV$_4$}
\label{sec: Cancel_subleading_MHV_4}
In this appendix we will do the explicit calculation of the subleading soft factor acting on the four-point MHV amplitude. The subleading soft theorem comes from the second term in the soft expansion
\begin{eqnarray}
\mathcal{M}_{n+1}^{(1)} & = & \sum_{a=1}^{n-1}\frac{[sa]\langle na\rangle^2}{\langle sa\rangle\langle ns\rangle^2}\left(\frac{\langle ns\rangle}{\langle na\rangle}\tilde{\lambda}_s\partial_{\tilde{\lambda}_a}+\frac{\langle as\rangle}{\langle an\rangle}\tilde{\lambda}_s\partial_{\tilde{\lambda}_n}\right)\mathcal{M}_n\cr
& = & \sum_{a=1}^{n-1}\left(\frac{[sa]\langle na\rangle}{\langle sa\rangle\langle ns\rangle}\tilde{\lambda}_s\partial_{\tilde{\lambda}_a}+\frac{[sa]\langle na\rangle}{\langle ns\rangle^2}\tilde{\lambda}_s\partial_{\tilde{\lambda}_n}\right)\mathcal{M}_n
\end{eqnarray}
The second term in this vanishes because we can use momentum conservation to write
\begin{equation}
\sum_{a=1}^{n-1}\frac{[sa]\langle na\rangle}{\langle ns\rangle^2}\tilde{\lambda}_s\partial_{\tilde{\lambda}_n}\mathcal{M}_n = \frac{1}{\langle ns\rangle^2}[s\left(-\sum_{a=1}^{n-1}a]\langle a\right)n\rangle \tilde{\lambda}_s\partial_{\tilde{\lambda}_n}\mathcal{M}_n = \frac{1}{\langle ns\rangle^2}[sn]\langle nn\rangle\tilde{\lambda}_s\partial_{\tilde{\lambda}_n}\mathcal{M}_n=0.
\end{equation}
Therefore
\begin{equation}
\mathcal{M}_{n+1}^{(1)} = \sum_{a=1}^{n-1}\frac{[sa]\langle na\rangle}{\langle sa\rangle\langle ns\rangle}\tilde{\lambda}_s\partial_{\tilde{\lambda}_a}\mathcal{M}_n.
\end{equation}
Using Mathematica, and simplifying with the Schouten identity we obtain
\begin{eqnarray}
\mathcal{M}_{5}^{(1)} & = & -\frac{\langle 12\rangle^6[35](\langle 35\rangle(\langle 14\rangle[15]+\langle 24\rangle[25]+\langle 34\rangle[35])+\langle 34\rangle\langle 45\rangle[45])}{\langle 13\rangle\langle 14\rangle\langle 23\rangle\langle 24\rangle\langle 34\rangle\langle 35\rangle\langle 45\rangle^2}\cr
& + & \frac{\langle 12\rangle^6[15][34](\langle 15\rangle(\langle 14\rangle[15]+\langle 24\rangle[25]+\langle 34\rangle[35])+\langle 14\rangle\langle 45\rangle[45])}{\langle 13\rangle\langle 15\rangle\langle 23\rangle\langle 24\rangle\langle 34\rangle\langle 45\rangle^2}\delta(1,4)\cr
& + & \frac{\langle 12\rangle^6[25][34](\langle 25\rangle(\langle 14\rangle[15]+\langle 24\rangle[25]+\langle 34\rangle[35])+\langle 24\rangle\langle 45\rangle[45])}{\langle 13\rangle\langle 14\rangle\langle 23\rangle\langle 25\rangle\langle 34\rangle\langle 45\rangle^2}\delta(2,4)\cr
& + & \frac{\langle 12\rangle^6\langle 13\rangle[15][34][35]}{\langle 14\rangle\langle 15\rangle\langle 23\rangle\langle 24\rangle\langle 34\rangle\langle 35\rangle}\delta(1,3) + \frac{\langle 12\rangle^6\langle 23\rangle[25][34][35]}{\langle 13\rangle\langle 14\rangle\langle 24\rangle\langle 25\rangle\langle 34\rangle\langle 35\rangle}\delta(2,3)\cr
& = & \frac{\langle 12\rangle^6[34][35](\langle 35\rangle(\langle 14\rangle[15]+\langle 24\rangle[25]+\langle 34\rangle[35])+\langle 34\rangle\langle 45\rangle[45])}{\langle 13\rangle\langle 14\rangle\langle 23\rangle\langle 24\rangle\langle 35\rangle\langle 45\rangle^2}\delta(3,4).
\end{eqnarray}
Now we want to impose momentum conservation. Since we have five particles, we need to impose the five point momentum conservation but where the fifth one is soft and weighted by $\epsilon$. We will use the following
\begin{equation}
\langle a 1\rangle[1b] + \langle a 2\rangle[2b]  + \langle a 3\rangle[3b] + \langle a 4\rangle[4b] + \epsilon\langle a 5\rangle[5b] = 0.
\end{equation}
When $a=4,b=5$ the last two terms are 0 so we obtain
\begin{eqnarray}
\mathcal{M}_{5}^{(1)} & = & -\frac{\langle 12\rangle^6[35][45]}{\langle 13\rangle\langle 14\rangle\langle 23\rangle\langle 24\rangle\langle 35\rangle\langle 45\rangle}\cr
& + & \frac{\langle 12\rangle^6[15][34][45]}{\langle 13\rangle\langle 15\rangle\langle 23\rangle\langle 24\rangle\langle 34\rangle\langle 45\rangle}{\color{red}\langle 14\rangle \delta(1,4)}+ \frac{\langle 12\rangle^6[25][34][45]}{\langle 13\rangle\langle 14\rangle\langle 23\rangle\langle 25\rangle\langle 34\rangle\langle 45\rangle}{\color{red}\langle 24\rangle \delta(2,4)}\cr
& + & \frac{\langle 12\rangle^6[15][34][35]}{\langle 14\rangle\langle 15\rangle\langle 23\rangle\langle 24\rangle\langle 34\rangle\langle 35\rangle}{\color{red}\langle 13\rangle \delta(1,3)} + \frac{\langle 12\rangle^6[25][34][35]}{\langle 13\rangle\langle 14\rangle\langle 24\rangle\langle 25\rangle\langle 34\rangle\langle 35\rangle}{\color{red}\langle 23\rangle \delta(2,3)}\cr
& + & \frac{\langle 12\rangle^6[35][45]}{\langle 13\rangle\langle 14\rangle\langle 23\rangle\langle 24\rangle\langle 45\rangle}{\color{red} [34]\langle 34\rangle \delta(3,4)}.
\end{eqnarray}
We see that all the distributional terms vanish because they all look like $z\delta(z)$. 

It is necessary to comment on the fact that in the general case in the main text we used the Schouten identity for the subleading case but here we have used momentum conservation. Momentum conservation suffices in the four point example because there are few enough particles to yield only the vanishing term upon imposing the constraint. For higher than four points, there will be more than one term after imposing momentum conservation and only one of them will vanish. Therefore, generally the momentum conservation argument does not suffice and one needs to use the above argument with the Schouten identity. 

\section{Full five-point MHV when $\epsilon=1$}
\label{sec: full_five_point}
In the case of the five-point function, beyond the subleading level there are no terms so if we let $\epsilon=1$, then the two terms should sum to give the full five-point MHV amplitude. Setting $\epsilon=1$ means, that we no longer have to care about the orders in $\epsilon$ because there is no longer an expansion being done. In order to show this, it will be useful to have everything written in terms of the four-point amplitude. The five-point amplitude is
\begin{equation}
\mathcal{M}_5 = \left[\frac{[53]\langle 23\rangle\langle 13\rangle}{\langle 53\rangle\langle 15\rangle\langle 25\rangle} +  \frac{[45][53]\langle 43\rangle}{\langle 45\rangle[43]\langle 53\rangle} + \frac{[45]\langle 14\rangle\langle 24\rangle}{\langle 45\rangle\langle 15\rangle\langle 25\rangle}\right]\mathcal{M}_4.
\end{equation}
We need to show that the following sum is equivalent to the full five-point amplitude
\begin{eqnarray}
\mathcal{M}_4^{(0)} + \mathcal{M}_4^{(1)} & = & \left[\frac{[51]\langle 14\rangle^2}{\langle 51\rangle\langle 54\rangle^2} + \frac{[52]\langle 24\rangle^2 }{\langle 52\rangle\langle 54\rangle^2} + \frac{[53]\langle 34\rangle^2}{\langle 53\rangle\langle 54\rangle^2}\right]\mathcal{M}_4 +  \frac{[45][53]\langle 43\rangle}{[43]\langle 45\rangle\langle 53\rangle}\mathcal{M}_4.
\end{eqnarray}
We see that this is equivalent to showing that the following is zero
\begin{small}
\begin{eqnarray}
X & = & \frac{[53]\langle 23\rangle\langle 13\rangle}{\langle 53\rangle\langle 15\rangle\langle 25\rangle}+ \frac{[45]\langle 14\rangle\langle 24\rangle}{\langle 45\rangle\langle 15\rangle\langle 25\rangle} - \frac{[51]\langle 14\rangle^2}{\langle 51\rangle\langle 54\rangle^2} - \frac{[52]\langle 24\rangle^2 }{\langle 52\rangle\langle 54\rangle^2} - \frac{[53]\langle 34\rangle^2}{\langle 53\rangle\langle 54\rangle^2} \cr
& = & \frac{\left[[53]\langle 23\rangle\langle 13\rangle\langle 54\rangle^2- [45]\langle 14\rangle\langle 24\rangle\langle 53\rangle\langle 54\rangle +[51]\langle 14\rangle^2\langle 53\rangle\langle 25\rangle + [52]\langle 24\rangle^2\langle 53\rangle\langle 15\rangle  - [53]\langle 34\rangle^2\langle 15\rangle\langle 25\rangle\right]}{\langle 53\rangle\langle 15\rangle\langle 25\rangle\langle 54\rangle^2}.
\end{eqnarray}
\end{small}Now we can look at the numerator and use momentum conservation
\begin{eqnarray}
X_{\rm{num}} & = & [53]\langle 23\rangle\langle 13\rangle\langle 54\rangle^2+ \left([52]\langle 21\rangle+[53]\langle 31\rangle\right)\langle 24\rangle\langle 53\rangle\langle 54\rangle - [53]\langle 34\rangle^2\langle 15\rangle\langle 25\rangle\cr
& - & \left([52]\langle 24\rangle+[53]\langle 34\rangle\right)\langle 14\rangle\langle 53\rangle\langle 25\rangle +  [52]\langle 24\rangle^2\langle 53\rangle\langle 15\rangle  \cr
& = & [53]\left(\langle 23\rangle\langle 13\rangle\langle 54\rangle^2 + \langle 31\rangle\langle 24\rangle\langle 53\rangle\langle 54\rangle- \langle 34\rangle^2\langle 15\rangle\langle 25\rangle-\langle 34\rangle\langle 14\rangle\langle 53\rangle\langle 25\rangle\right)\cr
& + &[52]\langle 24\rangle\langle 53\rangle{\color{red}\left(\langle 21\rangle\langle 54\rangle - \langle 14\rangle\langle 25\rangle +  \langle 24\rangle\langle 15\rangle \right)}.
\end{eqnarray}
The second line vanishes using the Schouten identity and we are only left with the first line
\begin{eqnarray}
X_{\rm{num}} & = &[53]\left[\langle 13\rangle\langle 54\rangle\left(\langle 23\rangle\langle 54\rangle - \langle 24\rangle\langle 53\rangle\right)- \langle 34\rangle\langle 25\rangle\left(\langle 34\rangle\langle 15\rangle+\langle 14\rangle\langle 53\rangle\right)\right]\cr
& = & [53]\left[\langle 13\rangle\langle 54\rangle\langle 25\rangle\langle 34\rangle- \langle 34\rangle\langle 25\rangle\langle 13\rangle\langle 54\rangle\right] = 0
\end{eqnarray}
which also vanishes using the Schouten identity. Therefore, we have shown that the leading and subleading terms give the full five-point amplitude in the $\epsilon=1$ limit. This is consistent with~\cite{Cachazo:2014fwa}.
\section{Simplification of distributional terms: Sub-subleading on MHV$_4$}
\label{sec: simplify_dist_MHV_4}
If we act on the four-point amplitude with the sub-subleading soft factor and account for all delta function terms coming from poles in the amplitude, we get the following from Mathematica
\begin{eqnarray}
\mathcal{M}_5^{(2)} & = & -\frac{2\langle 12\rangle^6[15][35][45]}{\langle 14\rangle\langle 23\rangle\langle 24\rangle\langle 34\rangle\langle 35\rangle}\delta(1,3)-\frac{2\langle 12\rangle^6[25][35][45]}{\langle 13\rangle\langle 14\rangle\langle 24\rangle\langle 34\rangle\langle 35\rangle}\delta(2,3)\cr
& + & \frac{2\langle 12\rangle^6[15][35][45]}{\langle 13\rangle\langle 23\rangle\langle 24\rangle\langle 34\rangle\langle 45\rangle}\delta(1,4) + \frac{2\langle 12\rangle^6[25][35][45]}{\langle 13\rangle\langle 14\rangle\langle 23\rangle\langle 34\rangle\langle 45\rangle}\delta(2,4)\cr
& + & \frac{2\langle 12\rangle^6[35][45](\langle 35\rangle[35]+\langle 45\rangle[45])}{\langle 13\rangle\langle 14\rangle\langle 23\rangle\langle 24\rangle\langle 35\rangle\langle 45\rangle}\delta(3,4)\cr
& + & \frac{\langle 12\rangle^6[15][34][35](\langle 15\rangle[15]+\langle 35\rangle[35])}{\langle 14\rangle\langle 15\rangle\langle 23\rangle\langle 24\rangle\langle 34\rangle\langle 35\rangle}\partial \delta(1,3)+\frac{\langle 12\rangle^6[25][34][35](\langle 25\rangle[25]+\langle 35\rangle[35])}{\langle 13\rangle\langle 14\rangle\langle 24\rangle\langle 25\rangle\langle 34\rangle\langle 35\rangle}\partial \delta(2,3)\cr
& + & \frac{\langle 12\rangle^6[15][34][45](\langle 15\rangle[15]+\langle 45\rangle[45])}{\langle 13\rangle\langle 15\rangle\langle 23\rangle\langle 24\rangle\langle 34\rangle\langle 45\rangle}\partial \delta(1,4) + \frac{\langle 12\rangle^6[25][34][45](\langle 25\rangle[25]+\langle 45\rangle[45])}{\langle 13\rangle\langle 14\rangle\langle 23\rangle\langle 25\rangle\langle 34\rangle\langle 45\rangle}\partial \delta(2,4)\cr
& + & \frac{\langle 12\rangle^6[34][35][45](\langle 35\rangle[35]+\langle 45\rangle[45])}{\langle 13\rangle\langle 14\rangle\langle 23\rangle\langle 24\rangle\langle 35\rangle\langle 45\rangle}\partial \delta(3,4)\cr
& - & \frac{2\langle 12\rangle^6[15][34][35][45]}{\langle 15\rangle\langle 23\rangle\langle 24\rangle\langle 34\rangle}\delta(1,3)\delta(1,4) + \frac{2\langle 12\rangle^6[25][34][35][45]}{\langle 13\rangle\langle 14\rangle\langle 24\rangle\langle 35\rangle}\delta(2,3)\delta(3,4)\cr
& + & \frac{2\langle 12\rangle^6[15][34][35][45]}{\langle 14\rangle\langle 23\rangle\langle 24\rangle\langle 35\rangle}\delta(1,3)\delta(3,4) - \frac{2\langle 12\rangle^6[25][34][35][45]}{\langle 13\rangle\langle 14\rangle\langle 23\rangle\langle 45\rangle}\delta(2,4)\delta(3,4)\cr
& - & \frac{2\langle 12\rangle^6[15][34][35][45]}{\langle 13\rangle\langle 23\rangle\langle 24\rangle\langle 45\rangle}\delta(1,4)\delta(3,4) - \frac{2\langle 12\rangle^6[25][34][35][45]}{\langle 13\rangle\langle 14\rangle\langle 25\rangle\langle 34\rangle}\delta(2,3)\delta(2,4)
\end{eqnarray}
We have three types of terms here that must be considered separately. There are terms that are proportional to a delta function, terms that are proportional to derivatives of delta functions and terms that proportional to products of delta functions. We expect the last to vanish and the second set to combine with the first. We first look at the terms containing products of delta functions, that we label by the subscript ``dd". In order to simplify them we will conveniently multiply the terms by 1, making sure to keep track of factors of epsilon that come along with $\lambda_5$.
\begin{eqnarray}
\mathcal{M}_{5,\rm{dd}}^{(2)} & = & - {\color{red}\frac{\epsilon\langle 52\rangle}{\epsilon\langle 52\rangle}}\frac{2\langle 12\rangle^6[15][34][35][45]}{\langle 15\rangle\langle 23\rangle\langle 24\rangle\langle 34\rangle}\delta(1,3)\delta(1,4) + {\color{red}\frac{\epsilon\langle 51\rangle}{\epsilon\langle 51\rangle}}\frac{2\langle 12\rangle^6[25][34][35][45]}{\langle 13\rangle\langle 14\rangle\langle 24\rangle\langle 35\rangle}\delta(2,3)\delta(3,4)\cr
& + & {\color{red}\frac{\epsilon\langle 52\rangle}{\epsilon\langle 52\rangle}}\frac{2\langle 12\rangle^6[15][34][35][45]}{\langle 14\rangle\langle 23\rangle\langle 24\rangle\langle 35\rangle}\delta(1,3)\delta(3,4) - {\color{red}\frac{\epsilon\langle 51\rangle}{\epsilon\langle 51\rangle}}\frac{2\langle 12\rangle^6[25][34][35][45]}{\langle 13\rangle\langle 14\rangle\langle 23\rangle\langle 45\rangle}\delta(2,4)\delta(3,4)\cr
& - & {\color{red}\frac{\epsilon\langle 52\rangle}{\epsilon\langle 52\rangle}}\frac{2\langle 12\rangle^6[15][34][35][45]}{\langle 13\rangle\langle 23\rangle\langle 24\rangle\langle 45\rangle}\delta(1,4)\delta(3,4) - {\color{red}\frac{\epsilon\langle 51\rangle}{\epsilon\langle 51\rangle}}\frac{2\langle 12\rangle^6[25][34][35][45]}{\langle 13\rangle\langle 14\rangle\langle 25\rangle\langle 34\rangle}\delta(2,3)\delta(2,4)\cr
& = & \frac{[34]}{\langle 34\rangle}\frac{2\langle 12\rangle^6\left([13]\langle 32\rangle+[14]\langle 42\rangle\right)[35][45]}{\langle 15\rangle\langle 23\rangle\langle 24\rangle\langle 52\rangle}\delta(1,3)\delta(1,4) \cr
& - &  \frac{[34]}{\langle 24\rangle}\frac{2\langle 12\rangle^6\left([23]\langle 31\rangle+[24]\langle 41\rangle\right)[35][45]}{\langle 13\rangle\langle 14\rangle\langle 35\rangle\langle 51\rangle}\delta(2,3)\delta(3,4)\cr
& - & \frac{[34]}{\langle 14\rangle}\frac{2\langle 12\rangle^6[15]\left([31]\langle 12\rangle+[34]\langle 42\rangle\right)[45]}{\langle 23\rangle\langle 24\rangle\langle 35\rangle\langle 52\rangle}\delta(1,3)\delta(3,4) \cr
& + &  \frac{[34]}{\langle 23\rangle}\frac{2\langle 12\rangle^6[25][35]\left([42]\langle 21\rangle+[43]\langle 31\rangle\right)}{\langle 13\rangle\langle 14\rangle\langle 45\rangle\langle 51\rangle}\delta(2,4)\delta(3,4)\cr
& + & \frac{[34]}{\langle 13\rangle}\frac{2\langle 12\rangle^6[15][35]\left([41]\langle 12\rangle+[43]\langle 32\rangle\right)}{\langle 23\rangle\langle 24\rangle\langle 45\rangle\langle 52\rangle}\delta(1,4)\delta(3,4) \cr
& + &  \frac{[34]}{\langle 34\rangle}\frac{2\langle 12\rangle^6\left([23]\langle 31\rangle+[24]\langle 41\rangle\right)[35][45]}{\langle 13\rangle\langle 14\rangle\langle 25\rangle\langle 51\rangle}\delta(2,3)\delta(2,4)
\end{eqnarray}
We see that each of these terms goes to 0 since they look like $z\delta(z)$. To ensure that this is, in fact true, we have pulled out the factor of $\frac{0}{0}$ in each term that does not vanish. We also note that this simplification was at the same order in $\epsilon$ as the sub-subleading expansion because the factors of $\epsilon$ cancelled between the numerator and the denominator. Now we can look at the terms that have derivatives of delta functions which, for notational convenience, we have written as $\partial \delta(a,b)$ and labeled by a subscript ``deltaderiv". These terms are
\begin{eqnarray}
\mathcal{M}_{5,\rm{deltaderiv}}^{(2)} & = & {\color{red}\frac{\langle 42\rangle}{\langle 42\rangle}}\frac{\langle 12\rangle^6[15][34][35](\langle 15\rangle[15]+\langle 35\rangle[35])}{\langle 14\rangle\langle 15\rangle\langle 23\rangle\langle 24\rangle\langle 34\rangle\langle 35\rangle}\partial \delta(1,3)\cr
& + & {\color{red}\frac{\langle 41\rangle}{\langle 41\rangle}}\frac{\langle 12\rangle^6[25][34][35](\langle 25\rangle[25]+\langle 35\rangle[35])}{\langle 13\rangle\langle 14\rangle\langle 24\rangle\langle 25\rangle\langle 34\rangle\langle 35\rangle}\partial \delta(2,3)\cr
& + & {\color{red}\frac{\langle 23\rangle}{\langle 23\rangle}}\frac{\langle 12\rangle^6[15][34][45](\langle 15\rangle[15]+\langle 45\rangle[45])}{\langle 13\rangle\langle 15\rangle\langle 23\rangle\langle 24\rangle\langle 34\rangle\langle 45\rangle}\partial \delta(1,4) \cr
& + & {\color{red}\frac{\langle 13\rangle}{\langle 13\rangle}}\frac{\langle 12\rangle^6[25][34][45](\langle 25\rangle[25]+\langle 45\rangle[45])}{\langle 13\rangle\langle 14\rangle\langle 23\rangle\langle 25\rangle\langle 34\rangle\langle 45\rangle}\partial \delta(2,4)\cr
& + & \frac{\langle 12\rangle^6[34][35][45](\langle 35\rangle[35]+\langle 45\rangle[45])}{\langle 13\rangle\langle 14\rangle\langle 23\rangle\langle 24\rangle\langle 35\rangle\langle 45\rangle}\partial \delta(3,4)
\end{eqnarray}
where we have added in the red factors of one in order to simplify them later. Note that none of the factors had any $\epsilon$s since they did not involve $\lambda_5$. Now we can use momentum conservation in the numerator
\begin{eqnarray}
\mathcal{M}_{5,\rm{deltaderiv}}^{(2)} & = & \frac{\langle 12\rangle^6[15]\left([31]\langle 12\rangle+\epsilon[35]\langle 52\rangle\right)[35](\langle 15\rangle[15]+\langle 35\rangle[35])}{\langle 14\rangle\langle 15\rangle\langle 23\rangle\langle 24\rangle^2\langle 34\rangle\langle 35\rangle}\partial \delta(1,3)\cr
& + & \frac{\langle 12\rangle^6[25]\left([32]\langle 21\rangle+\epsilon[35]\langle 51\rangle\right)[35](\langle 25\rangle[25]+\langle 35\rangle[35])}{\langle 13\rangle\langle 14\rangle^2\langle 24\rangle\langle 25\rangle\langle 34\rangle\langle 35\rangle}\partial \delta(2,3)\cr
& - & \frac{\langle 12\rangle^6[15]\left(\langle 21\rangle[14]+\epsilon\langle 25\rangle[54]\right)[45](\langle 15\rangle[15]+\langle 45\rangle[45])}{\langle 13\rangle\langle 15\rangle\langle 23\rangle^2\langle 24\rangle\langle 34\rangle\langle 45\rangle}\partial \delta(1,4) \cr
& - & \frac{\langle 12\rangle^6[25]\left(\langle 12\rangle[24]+\epsilon\langle 15\rangle[54]\right)[45](\langle 25\rangle[25]+\langle 45\rangle[45])}{\langle 13\rangle^2\langle 14\rangle\langle 23\rangle\langle 25\rangle\langle 34\rangle\langle 45\rangle}\partial \delta(2,4)\cr
& + & \frac{\langle 12\rangle^6[34][35][45](\langle 35\rangle[35]+\langle 45\rangle[45])}{\langle 13\rangle\langle 14\rangle\langle 23\rangle\langle 24\rangle\langle 35\rangle\langle 45\rangle}\partial \delta(3,4)
\end{eqnarray}
We see that some of the terms go like $\epsilon$ which means they will not contribute to the $\mathcal{O}(\epsilon^{-1})$ order in the soft expansion. Therefore, we can discard those terms
\begin{eqnarray}
\mathcal{M}_{5,\rm{deltaderiv}}^{(2)} & = & \frac{\langle 12\rangle^7[15][35](\langle 15\rangle[15]+\langle 35\rangle[35])}{\langle 14\rangle\langle 15\rangle\langle 23\rangle\langle 24\rangle^2\langle 34\rangle\langle 35\rangle}\delta(1,3)\cr
& - & \frac{\langle 12\rangle^7[25][35](\langle 25\rangle[25]+\langle 35\rangle[35])}{\langle 13\rangle\langle 14\rangle^2\langle 24\rangle\langle 25\rangle\langle 34\rangle\langle 35\rangle}\delta(2,3) \cr
& - & \frac{\langle 12\rangle^7[15][45](\langle 15\rangle[15]+\langle 45\rangle[45])}{\langle 13\rangle\langle 15\rangle\langle 23\rangle^2\langle 24\rangle\langle 34\rangle\langle 45\rangle}\delta(1,4) \cr
& + & \frac{\langle 12\rangle^7[25][45](\langle 25\rangle[25]+\langle 45\rangle[45])}{\langle 13\rangle^2\langle 14\rangle\langle 23\rangle\langle 25\rangle\langle 34\rangle\langle 45\rangle}\delta(2,4)\cr
& - & \frac{\langle 12\rangle^6[35][45](\langle 35\rangle[35]+\langle 45\rangle[45])}{\langle 13\rangle\langle 14\rangle\langle 23\rangle\langle 24\rangle\langle 35\rangle\langle 45\rangle} \delta(3,4)
\end{eqnarray}
We removed the terms that were higher order in $\epsilon$ since they do not contribute to the sub-subleading soft expansion and then used the fact that $x\,\partial\delta(x) = -\delta(x)$. Now we can add these to the single delta function terms 
\begin{eqnarray}
\mathcal{M}_5^{(2)} & = & -\frac{\langle 12\rangle^6[15][35][45]}{\langle 14\rangle\langle 23\rangle\langle 24\rangle\langle 34\rangle\langle 35\rangle}\delta(1,3) -  \frac{\langle 12\rangle^6[25][35] [45]}{\langle 13\rangle\langle 14\rangle\langle 24\rangle\langle 34\rangle\langle 35\rangle}\delta(2,3) +\frac{\langle 12\rangle^6[15][45][35]}{\langle 13\rangle\langle 23\rangle\langle 24\rangle\langle 34\rangle\langle 45\rangle}\delta(1,4) \cr
& + &  \frac{\langle 12\rangle^6[25][45][35]}{\langle 13\rangle\langle 14\rangle\langle 23\rangle\langle 34\rangle\langle 45\rangle}\delta(2,4) + \frac{\langle 12\rangle^6[35][45](\langle 35\rangle[35]+\langle 45\rangle[45])}{\langle 13\rangle\langle 14\rangle\langle 23\rangle\langle 24\rangle\langle 35\rangle\langle 45\rangle}\delta(3,4) 
\end{eqnarray}

\section{Equivalence of MHV Graviton Formulas}\label{app:equiv}

In this appendix we prove the equivalence between the formula we use for the MHV graviton amplitude and Hodges' formula. Recall that Hodges' formula states that the $n$-point MHV graviton amplitude is given by 
\begin{equation}
M_n = \frac{\langle 12\rangle^6}{\langle 23\rangle^2\langle 31\rangle^2}\mbox{det}\Phi_{123}^{123} = \frac{\langle 12\rangle^6}{\langle 23\rangle^2\langle 31\rangle^2}\sum_{\{i_4,\cdots,i_n\}\in\{4,\cdots,n\}}\varepsilon_{i_4\cdots i_n}\Phi_{4i_4}\cdots\Phi_{ni_n}
\end{equation}
where 
\begin{equation}
\Phi_{ab} = \frac{[ab]}{\langle ab\rangle}, \ a\neq b, \ \ \Phi_{ab} = -\sum_{c=1; c\neq a}^n\frac{[ac]}{\langle ac\rangle}\frac{\langle 1c\rangle\langle 2c\rangle}{\langle 1a\rangle\langle 2a\rangle}, \ a=b.
\end{equation}
The proposition is that we can alternatively write the $n$-point MHV graviton amplitude as
\begin{equation}
M_n = \mathcal{F}_n\left(\prod_{i=1,j>i}^n\frac{1}{\langle ij\rangle}\right)
\end{equation}
where $\mathcal{F}_n$ is a function that has no poles. We will show that this is equivalent by first demonstrating that from Hodges' formula, any term will have a pole of maximum order 2. We will then show that the terms having an order 2 pole will cancel each other. Since only order 1 poles remain, the common denominator is the product of all single poles.

First, let us look at the orders of poles appearing in Hodges' formula. The cleanest way to do this is to consider one pair $(i,j)$ and look at all the terms that have $\langle ij\rangle$ in the denominator more than once. One way to get two factors in the denominator is $\Phi_{ij}\Phi_{ji} = \frac{[ij]^2}{\langle ij\rangle^2}$. This will be multiplied by $\Phi_{ab}$'s where $a,b\neq i,j$ and therefore there will be no more factors of $\frac{1}{\langle ij\rangle}$ in this term. We can never have $\Phi_{ij}$ multiplied by $\Phi_{ii}$ or $\Phi_{jj}$ so the only other option is when we have $\Phi_{ii}\Phi_{jj}$. If we write this out
\begin{equation}
\Phi_{ii}\Phi_{jj} = \sum_{c=1;c\neq i}^n\frac{[ic]}{\langle ic\rangle}\frac{\langle 1c\rangle\langle 2c\rangle}{\langle 1i\rangle\langle 2i\rangle}\sum_{b=1;b\neq j}^n\frac{[jb]}{\langle jb\rangle}\frac{\langle 1b\rangle\langle 2b\rangle}{\langle 1j\rangle\langle 2j\rangle}
\end{equation}
we see that only when $c=j$ and $b=i$ we will get the second order pole we care about. Writing out just that term in the product gives 
\begin{equation}
(\Phi_{ii}\Phi_{jj})_{\langle ij\rangle^{-2}} = \frac{[ij]^2}{\langle ij\rangle^2}
\end{equation}
which is exactly equal to $\Phi_{ij}\Phi_{ji}$. Terms containing the first instance of the double pole will be of the form
\begin{equation}
{M_n}_{\{(i,j),(j,i)\}} = \frac{\langle 12\rangle^6}{\langle 23\rangle^2\langle 31\rangle^2}\sum_{i_k\in\{4,\cdots,n\}/\{i,j\}}\varepsilon_{i_4\cdots i_i=j\cdots i_j=i\cdots i_n}\Phi_{4i_4}\cdots \Phi_{ij}\cdots\Phi_{ji}\cdots\Phi_{ni_n}.
\end{equation}
Terms containing the second instance of the double pole will be of the form
\begin{equation}
{M_n}_{\{(i,i),(j,j)\}} = \frac{\langle 12\rangle^6}{\langle 23\rangle^2\langle 31\rangle^2}\sum_{i_k\in\{4,\cdots,n\}/\{i,j\}}\varepsilon_{i_4\cdots i_i=i\cdots i_j=j\cdots i_n}\Phi_{4i_4}\cdots \Phi_{ii}\cdots\Phi_{jj}\cdots\Phi_{ni_n}.
\end{equation}
Summing these together, we see that because $\varepsilon$ is totally antisymmetric, these two terms will have a relative minus sign with respect to each other and the double pole terms will exactly cancel. The only remaining double poles are the ones in the prefactor $\langle 23\rangle^2\langle 31\rangle^2$. The first term in the sum for $\Phi_{ab}$ when $a=b$, will contain $\langle 13\rangle\langle 23\rangle$ and cancel one factor. 

Defining $\Phi_{ab}'=\Phi_{ab}$ for $a\neq b$ and $\Phi_{ab}' = \sum_{c=4;\neq a}^n\frac{[ac]}{\langle ac\rangle}\frac{\langle 1c\rangle\langle 2c\rangle}{\langle 1a\rangle\langle 2a\rangle}$ for $a=b$, we would like to show that 
\begin{equation}
\mbox{det}\Phi' = \sum_{\{i_4,\cdots,i_n\}\in\{4,\cdots,n\}}\varepsilon_{i_4\cdots i_n}\Phi_{4i_4}'\cdots\Phi_{ni_n}'=0
\end{equation}
To do this we will look at the structure of $\Phi'$ in more detail. We see that we can write it as 
\begin{equation}
\Phi' = \sum_{i=4,j>i}^n\frac{[ij]}{\langle ij\rangle}V_{ij}
\end{equation}
where the only non-zero elements of the matrices $V_{ij}$ are 
\begin{eqnarray}
(V_{ij})_{i-3,i-3} =  -\frac{\langle 1j\rangle\langle 2j\rangle}{\langle 1i\rangle\langle 2i\rangle}, \ \ (V_{ij})_{j-3,j-3} =  -\frac{\langle 1i\rangle\langle 2i\rangle}{\langle 1j\rangle\langle 2j\rangle}, \ \ (V_{ij})_{i-3,j-3} = (V_{ij})_{j-3,i-3} = 1.
\end{eqnarray}
Further, we note that $V_{ij} = -\frac{\langle 1i\rangle\langle 2i\rangle}{\langle 1j\rangle\langle 2j\rangle}v_{ij}\otimes v_{ij}^T$ where $v_{ij} = -\frac{\langle 1j\rangle\langle 2j\rangle}{\langle 1i\rangle\langle 2i\rangle}\hat{e}_{i-3}+\hat{e}_{j-3}$ and $\hat{e}_k$ are the standard basis vectors for an $n-3$ dimensional space. There are $\frac{(n-3)(n-4)}{2}$ of the vectors $v_{ij}$. If less than $n-3$ of them are linearly independent, then the determinant of the matrix $\Phi'$ will be 0. Obviously, $v_{ij}$ and $v_{k\ell}$ for $i,j\neq k,\ell$ will be independent. We can also show that each triplet $v_{ij},v_{jk},v_{ik}$ are not linearly independent. If they are linearly independent then there will not exist non-zero scalars $\alpha,\beta,\gamma$ such that $\alpha v_{ij}+\beta v_{jk}+\gamma v_{ik}=0$. If we write this out explicitly
\begin{eqnarray}
\alpha \left[-\frac{\langle 1j\rangle\langle 2j\rangle}{\langle 1i\rangle\langle 2i\rangle}\hat{e}_{i-3}+\hat{e}_{j-3}\right] + \beta \left[-\frac{\langle 1k\rangle\langle 2k\rangle}{\langle 1j\rangle\langle 2j\rangle}\hat{e}_{j-3}+\hat{e}_{k-3}\right] + \gamma \left[-\frac{\langle 1k\rangle\langle 2k\rangle}{\langle 1i\rangle\langle 2i\rangle}\hat{e}_{i-3}+\hat{e}_{k-3}\right] = 0.
\end{eqnarray}
The coefficient of each independent basis vector needs to be 0 independently for this to be true
\begin{eqnarray}
-\alpha\frac{\langle 1j\rangle\langle 2j\rangle}{\langle 1i\rangle\langle 2i\rangle}-\gamma\frac{\langle 1k\rangle\langle 2k\rangle}{\langle 1i\rangle\langle 2i\rangle} & = & 0 \cr
\alpha -\beta \frac{\langle 1k\rangle\langle 2k\rangle}{\langle 1j\rangle\langle 2j\rangle} & = & 0\rightarrow \alpha = \beta \frac{\langle 1k\rangle\langle 2k\rangle}{\langle 1j\rangle\langle 2j\rangle}\cr
\beta +\gamma & = & 0\rightarrow \gamma = -\beta.
\end{eqnarray}
Substituting for $\alpha$ and $\gamma$ in the first equation
\begin{equation}
\beta\left[\frac{\langle 1k\rangle\langle 2k\rangle}{\langle 1i\rangle\langle 2i\rangle}- \frac{\langle 1k\rangle\langle 2k\rangle}{\langle 1j\rangle\langle 2j\rangle}\frac{\langle 1j\rangle\langle 2j\rangle}{\langle 1i\rangle\langle 2i\rangle}\right] = 0
\end{equation}
we see that any choice of $\beta$ will work because the term in the parenthesis is always 0. Therefore, there exist non-zero scalars $\alpha,\beta,\gamma$ such that this linear combination of vectors is 0 and they are linearly dependent. 

For a given $n$ the set of vectors is 
\begin{equation}
\{v_{45},\cdots v_{4n},v_{56},\cdots, v_{5n}, \cdots v_{n-1,n}\}.
\end{equation}
Since $\{v_{n-2,n-1}, v_{n-2,n}, v_{n-1,n}\}$ are linearly dependent, we can remove $v_{n-1,n}$ from this set. Then since $\{v_{n-3,n-2},v_{n-3,n-1},v_{n-2,n-1}\}$ are linearly dependent, we can remove $v_{n-2,n-1}$. We can keep going in this manner. Since $\{v_{n-3,n-2},v_{n-3,n}, v_{n-2,n}\}$ are linearly dependent, we can remove $v_{n-2,n}$. In this way we can systematically remove vectors until we see that the only set of linearly independent vectors is 
\begin{equation}
\{v_{45},\cdots, v_{4n}\}.
\end{equation}
Since this is a set of $n-4$ vectors that are linearly independent, we see that the $n-3\times n-3$ matrix $\Phi'$ can be written as a sum over projectors and since only $n-4$ of them are linearly independent, the determinant of $\Phi'$ is always 0. 

\bibliographystyle{utphys}
\bibliography{cpb}

\end{document}